\pdfoutput=1
\documentclass[10pt, conference]{IEEEtran}
\ifCLASSINFOpdf
\else
\fi

\usepackage{caption}
\usepackage[caption=false,font=footnotesize]{subfig}
\usepackage{fixltx2e}
\usepackage[bookmarks=false]{hyperref}
\usepackage{xspace}
\usepackage{array}
\usepackage{graphicx}
\usepackage{color}
\usepackage{float}
\usepackage{listings}
\usepackage{flushend}

\usepackage{paralist}
\usepackage[british]{babel}
\usepackage[activate={true,nocompatibility},final,tracking=true,kerning=true,spacing=true,factor=1100,stretch=10,shrink=20]{microtype}

\newfloat{lstfloat}{htbp}{lop}
\floatname{lstfloat}{Listing}

\lstdefinelanguage{Gift}{}
\newcommand{\BLU}[1]{#1}

\newcommand{\code}[1]{\texttt{#1}}
\newcommand{\textq}[1]{{\sffamily \small #1}}
\newcommand{\spencerurl}{http://spencer-t.racing}

\newcommand{\sourcecodeurl}[1]{\spencerurl/source/test/#1}
\newcommand{\sourcecodelink}[1]{\href{\sourcecodeurl{#1}}{\code{#1}}}

\newcommand{\dataset}[1]{\href{\spencerurl/datasets/#1}{\code{#1}}}

\newcommand{\lquery}[1]{\href{\spencerurl/query/test/#1}{\textq{#1}}}
\newcommand{\abbrevlquery}[2]{\href{\spencerurl/query/test/#2}{\textq{#1}}}
\newcommand{\uquery}[1]{\url{\spencerurl/query/test/#1}}
\newcommand{\hoquery}[2]{\href{\spencerurl/query/test/#1(MutableObj())}{\textq{#1($#2$)}}}

\begin{document}
%
\title{Spencer: Interactive Heap Analysis for the Masses}


\author{\IEEEauthorblockN{Stephan Brandauer}
\IEEEauthorblockA{Uppsala University, Sweden\\
  \href{mailto:stephan.brandauer@it.uu.se}{\texttt{stephan.brandauer@it.uu.se}}}
\and
\IEEEauthorblockN{Tobias Wrigstad}
\IEEEauthorblockA{Uppsala University, Sweden\\
  \href{mailto:tobias.wrigstad@it.uu.se}{\texttt{tobias.wrigstad@it.uu.se}}}
}



%


\maketitle
\pagestyle{plain}

\begin{abstract}
  Programming language-design and run-time-implementation require detailed
  knowledge about the programs that users want to implement. Acquiring this
  knowledge is hard, and there is little tool support to effectively estimate
  whether a proposed tradeoff actually makes sense in the context of real world
  applications.

  Ideally, knowledge about behaviour of ``typical'' programs is 1) easily 
  obtainable, 2) easily reproducible, and 3) easily sharable.
  
  We present \emph{Spencer}, an open source web service and API framework for
  dynamic analysis of a continuously growing set of traces of standard program
  corpora. Users do not obtain traces on their own, but can instead send queries
  to the web service that will be executed on a set of program traces. Queries
  are built in terms of a set of query combinators that present a high level
  interface for working with trace data. Since the framework is high level, and
  there is a hosted collection of recorded traces, queries are easy to
  implement. Since the data sets are shared by the research community, results
  are reproducible. Since the actual queries run on one (or many) servers that
  provide analysis as a service, obtaining results is possible on commodity
  hardware.
  
  Data in Spencer is meant to be obtained once, and analysed often, making the
  overhead of data collection mostly irrelevant. This allows Spencer to collect
  more data than traditional tracing tools can afford within their performance
  budget. Results in Spencer are cached, making complicated analyses that build
  on cached primitive queries speedy.
\end{abstract}


%
\IEEEpeerreviewmaketitle

\section{Introduction}

Standardised program corpora are commonly used to evaluate research on run-time-
and compiler optimisations -- an optimisation gets implemented, and a program
corpus is used to demonstrate its merit. Similarly, language abstractions and
novel type systems often use the same corpora: programs from the corpus are
annotated, and if it is possible to get most of the code to compile without
large structural changes, this validates the utility of the type system
\cite{haller2010,westbrook2012}. This work makes these corpora available
\emph{earlier in the research process}: we want researchers to be able to know
whether -- or not -- the case they are optimising for exists in common programs,
beyond artificial examples they have in mind. To this end, we run program
corpora with comprehensive tracing, then preprocess the traces, and finally make
the data available to be queried using a web interface (and an API that serves
data as JSON formatted objects). Researchers can now implement queries that test
whether -- for instance -- an optimisation they have in mind actually optimises
a pattern that common programs encounter frequently.

Compared to traditional dynamic analysis, Spencer's approach strikes a
\emph{novel tradeoff}: Spencer caters explicitly to the use case where someone
wants to know something about ``common programs'', not something about a
specific program of their own. This means that Spencer's work flow is simpler
than that of traditional tools: a user does not need to locally run any
expensive analysis and can immediately start to work on analysing data. We hope
that making program analysis \emph{easier} will increase the chances that it is
done, and improve how thoroughly it is done before research progresses. The
tradeoff is that users can't easily run analyses on their own programs (they
need to run Spencer locally, or work with the Spencer developers to have their
data set added in this case).

Spencer is the result of scratching our own itch. As researchers in programming
languages and programming language designers, we are often in search of data
that can confirm or disprove hypotheses, or influence design decisions.
Commonly, we end up mining existing code bases for answers, or hints at answers,
which is error-prone and scales poorly. Spencer allows us to interactively
explore (so far, Java) programs to find input to design processes, to gauge
usefulness of designs and uncover their pain points.

While dynamic analysis oftentimes cannot produce a \emph{sound} answer (because
it is based on some particular runs of some particular programs for some
particular data), the answers it gives still provide guidance and anectodal
evidence that trancedes ``gut feeling'' and ``folklore truths''. The easy access
to quantifiable data has changed our way of approaching programming language
design.

\subsection{Contributions}

\begin{itemize}[--]
\item We present Spencer, a web service to query program traces and visualise
  results interactively. Spencer's focus is on \emph{heap analysis}: tracing
  connections between objects, studying individual objects and groups of objects
  throughout their lifetime, and uncovering useful invariant properties such as
  \emph{uniqueness}, \emph{immutability} and \emph{reachability} and tying these
  invariants to static properties such as classes or source locations as well as
  dynamic properties like the span of time an object was in use.
\item We introduce a query language that can be used to query the Spencer data
  set. Queries can easily be embedded in papers to allow readers to obtain an
  interactive result that is open to further exploration and refinement.
\item To demonstrate its usefulness, we implement several small case studies in
  Spencer and show the range of queries that it can support.
\item Since Spencer is a web service, traces are recorded once and analysed
  often. This fact amortises the tracing overhead and makes tracing a more
  comprehensive data set (including the standard library) reasonable.
\end{itemize}

\section{Tracing as a Service}

\subsection{Design Goals and Tradeoffs}

Spencer's design goals are making knowledge of ``typ\-ical'' program
behaviour easy to \emph{obtain}, to \emph{reproduce}, and to \emph{share}.

Spencer makes one big tradeoff to meet these goals: data in Spencer is not
provided by the user. Instead, the maintainers of the tool upload datasets that
are deemed, by the Spencer developers, to represent a wide range of application
domains, and users query those. This rules out uses of Spencer as a tool for
analysing custom programs, which precludes its use as a bug-finding tool. On the
other hand, all Spencer code including the tracing infrastructure is freely
available as open source so nothing stops a user from running their own local
Spencer service on their own data sets.

In the context of the design goals, we argue that this tradeoff is well
justified. The remainder of this section will explain how.

\subsubsection{Easily Obtainable Knowledge}

Putting the data in a hosted repository is what makes it possible to host the
tool online. Therefore, this makes knowledge available \emph{without setting up
  any tool chain or configuring any tools}.

Since datasets, once uploaded, never change, query results 
can be cached. Running the same query again, on its own, or as a subquery of larger queries, results can be fetched from the result
cache, rather than computed again. \emph{As some datasets are large, caching is
  fundamentally important}. As caching also speeds up similar queries (as
subexpressions of a query might be already in the cache), this mechanism is also
important for \emph{exploring data sets}: when exploring a dataset, most often a
query is modified step by step. This means that the sequence of queries a user
looks at commonly share subexpressions -- and these are cached. Anecdotally, a
query that selects all immutable objects from the \code{pmd} benchmark takes 65
seconds if it has never been computed before, 500--800ms if it has been computed
before by any user. The speedup is between 80$\times$--130$\times$.

%

\subsubsection{Easily Reproducible Knowledge}

Comparing the results of dynamic analyses can be tricky: different tools
implemented by different researchers for different purposes commonly focus on
tracing just those pieces of information those researchers need. Comparing the
output of these tools can therefore be hard. One example is that most tracing
tools do not record variable accesses -- they are so numerous that the overhead
of logging them is often deemed too high.

Spencer makes its data available to the public and hosts very comprehensive
datasets: since program traces in Spencer are produced only once, but analysed
often (with caching), the overhead of tracing becomes unimportant. This means
that Spencer aims to record ``everything'' that happens. Variable loads and
stores, method calls and exits, and field accesses -- all with a range of meta
information, such as access times, field names, calling objects and methods. We
hope that this wide range of information will allow different researchers and
collaborators to compare each other's results reliably.

\subsubsection{Easily Sharable Knowledge}

In Spencer, every query is expressed as a URL. A query that returns all mutable
objects in the dataset called ``test'' is expressed by this URL:

\begin{center}
\code{\uquery{MutableObj()}}.
\end{center}

Naturally, queries can be much more complex, leading to longer URLs.

\begin{figure}
  \centering
  \includegraphics[width=9cm]{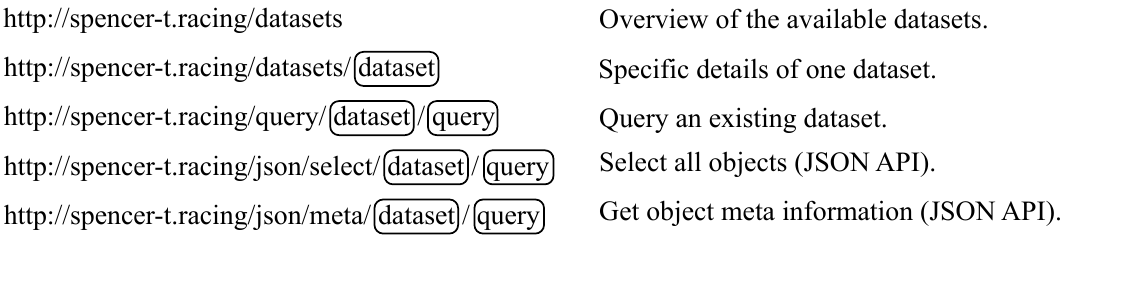}
  \caption{The available URLs that users can interact with.}\label{fig:urls}
\end{figure}

Since queries are URLs, \emph{sharing research results is just a matter of
  sharing a link}; and since results are cached on the server, sharing is
efficient. Because results are interactive, they serve as starting points for
exploration. A reader that wants to dig deeper to verify or dispute a hypothesis
that is not immediately addressed by the paper in which it appears may
experiment with adding or removing one particular data set from the results, or
check what objects cause a certain outlier, etc. This also makes it harder to
skew results by omitting data.

\subsection{The User Interface and Usability}

Spencer is a web service that lets users enter queries which will select sets of
objects (explained in more detail in Sec.~\ref{sec:queries}) and see information
about these selected objects. For the user, this means that they can use the
tool without any installation process or even downloads of data. For developers,
this makes it easy to add new visualisations, new data sets, or new primitive
queries. 

The user interface aims to be self documenting and the landing page,
\code{\url{\spencerurl}}, presents links to example queries in a tutorial style.
Figure~\ref{fig:urls} gives a brief overview of the sub pages that are available.


\subsubsection{Visualising Selections}

The ability to select objects (as covered in Section~\ref{sec:queries}) alone is
not useful for analysis of program traces -- these objects have to be tagged
with meta information, and this information needs to be visualised for a user.
Spencer provides a growing set of object variables:

\begin{center}
\begin{tabular}{l|p{4cm}l|c}
Name                   & Description                      \\\hline
\code{klass}           & Class of an object.              \\
\code{allocationSite}  & Allocation site (file, line).    \\
\code{thread}          & Allocating thread.               \\
\code{firstusage}      & Allocation time.                 \\
\code{lastusage}       & Last field access time.          \\
\code{lifeTime}        & Duration from allocation to \code{lastUsage}. \\
\end{tabular}
\end{center}
  
The object variables are used to visualise selection results.
For instance, the classes are visualised for a selection of objects in the form
of a bar chart that shows how many objects were created from a certain class. Spencer
distinguishes between categorical and numerical variables and the user interface
picks visualisations accordingly.

\begin{figure}
  \centering
  \subfloat[The classes and number of instances (zoomed in).]{
    \hspace{7mm}\includegraphics[trim=45 335 0 95, width=6.5cm, clip]{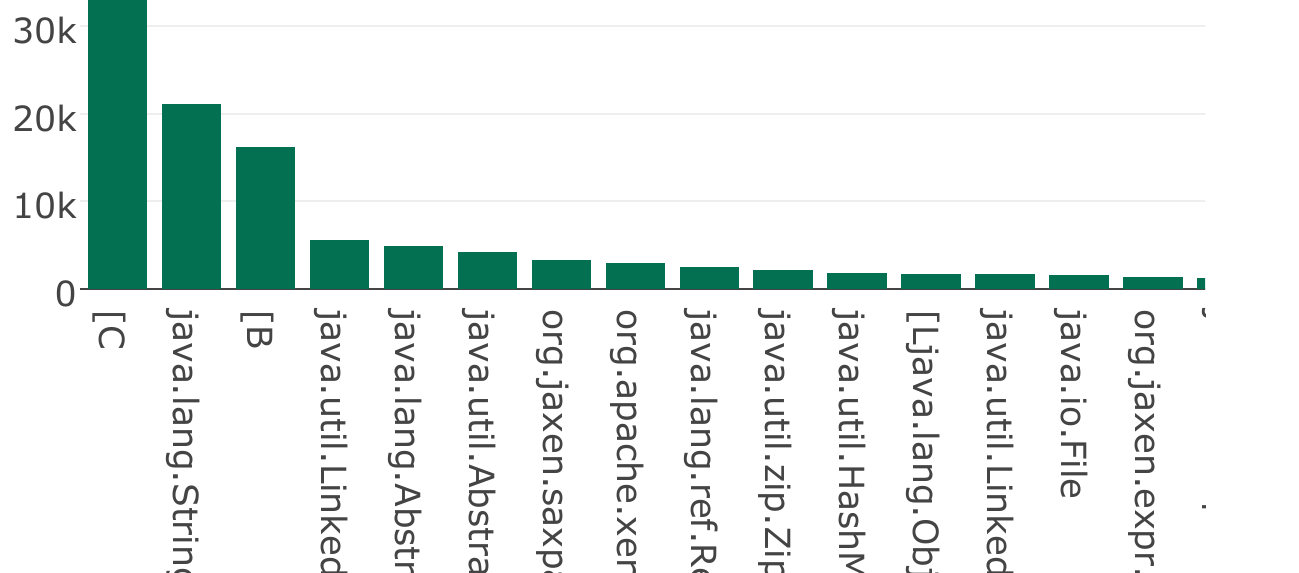}}\label{fig:ObjClassAbsPlot}
  \subfloat[The allocation sites (zoomed in).]{
    \includegraphics[trim=28 300 82 75, width=6.5cm, clip]{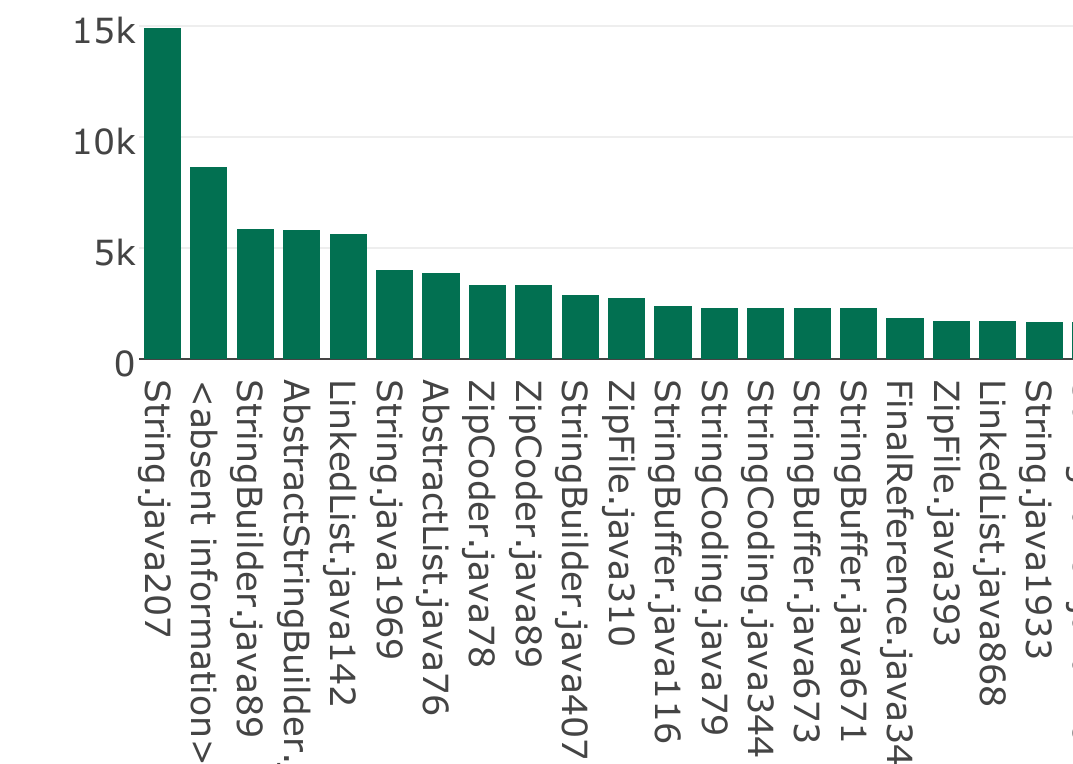}}\label{fig:ObjAllocationSiteAbsPlot}

  \subfloat[The variable \code{log10(lifeTime)}.]{
    \includegraphics[trim=28 0 0 0, width=5.2cm, clip]{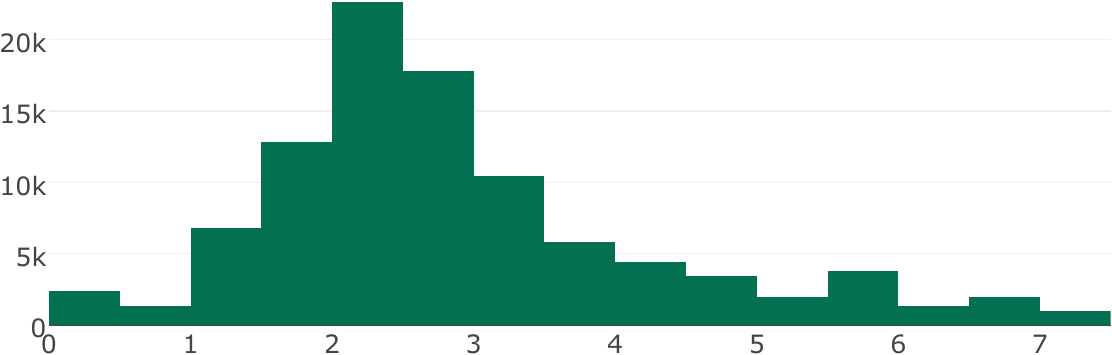}}\label{fig:ObjClassAbsPlot}
  \subfloat[The variable \code{log10(lifeTime)} (box plot).]{
    \hspace{-3mm}\includegraphics[trim=35 0 0 0, width=5cm, clip]{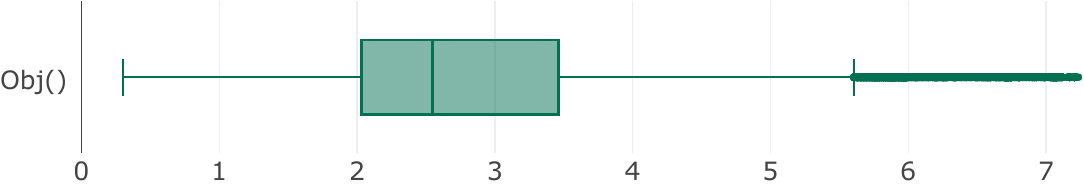}}\label{fig:ObjAllocationSiteAbsPlot}
  \caption{The query \lquery{Obj()} selects all objects. Figures (a) and (b)
    show visualisations of the categorical variables \code{klass} and
    \code{allocationSite}. Subfigures (c) and (d) show two visualisations for
    the numerical variable \code{log10(lifeTime)}. The lifetimes of objects
      appears to be a uni-modal distribution, with most objects being very short
      lived and few objects live very long.}\label{fig:ObjPlot}
\end{figure}

Figure~\ref{fig:ObjPlot}(a) shows an example, summarising the classes of
the objects selected by the query \lquery{Obj()} (in other words: all objects).

\section{Selection using Queries}\label{sec:queries}

Analyses in Spencer are written as high level queries. A query is a selection:
it returns as its result a set of object IDs (a unique integral value that
identifies each object).

The fact that there is a high level language of queries means that caching can
be effective: if queries in Spencer would be written, for instance, in a much
more expressive programming language, then many different programs could express
the same query. The cache system, however, would not be able to prove that
differently phrased (but equivalent) implementations of a query are in fact
equivalent. Caching would therefore speed up much fewer queries. The tradeoff
the design with the high level query language makes is that the queries a user
can run must be supported by the system explicitly -- it is possible that
Spencer does not support a certain query from being run. If a query algorithm
can not be expressed in Spencer, users can contribute the algorithm to the open
source service and thereby make it available for other users as well.

In Spencer, there are primitive queries, a set of basic selections that the
backend implements, and query combinators that users can use to combine queries
into more fine grained selections. Table~\ref{tab:queries} shows an overview of
the available queries. For example, the query \lquery{MutableObj()} returns a
set of object IDs that were mutated during a particular program's run. The query
\hoquery{ReachableFrom}{q} returns the set of objects that \emph{are reachable}
(via the heap, or via stack variables) from any object returned by the query
$q$. The query \hoquery{CanReach}{q} returns all objects that \emph{reach} an
object returned by $q$. Combined, the query \lquery{CanReach(ImmutableObj())}
returns all objects that are ``indirectly mutated'', meaning all objects that
are either mutated themselves, or objects that have fields referring to
indirectly mutated objects. The variants \textq{HeapReachableFrom($q$)} and
\textq{CanHeapReach($q$)} only consider reachability through fields,
\emph{i.e.,} it excludes stack variables.

The \hoquery{Deeply}{q} selects all objects that are
\emph{dominated} by the objects selected by $q$ in the object
graph. For example, if $o$ is in $q$, and $o'$ is an object which
can only be reached from $o$ (directly or indirectly), and $o''$
is an object which can be reached from $o$ by also from outside of
$o$, then $o'$ will be selected by \hoquery{Deeply}{q} but not
$o''$. 
Similarly, \hoquery{HeapDeeply}{q}
only considers fields, not stack variables. Sec.~\ref{sec:case-B} has an
example for its usage.

\begin{table}
  \begin{tabular}{lp{4.5cm}}
    \hline\hline
    Primitive Query                          & Selects all objects that are$\ldots$\\\hline
    \lquery{Obj()}                           & Any object. \\
    \lquery{MutableObj()}                    & mutated outside the constructor.\\
    \lquery{ImmutableObj()}                  & not mutated outside the constructor.\\
    \lquery{InstanceOf($c$)}                 & instances of the given class $c$.\\
    \lquery{StationaryObj()}                 & never written to after being read the first time.\\
    \lquery{TinyObj()}                       & never referring to any other objects.\\
    \lquery{UniqueObj()}                     & never aliased.\\
    \lquery{HeapUniqueObj()}                 & referred to from two fields at the same time.\\
    \hline \hline
    Query Combinator                         & Selects all objects that are$\ldots$\\
    \hline
    \hoquery{RefersTo}{q}                    & All objects that ever have a variable or field referring to an object in $q$.\\
    \hoquery{HeapRefersTo}{q}                & All objects that ever have a field referring to an object in $q$.\\
    \hoquery{ReferredFrom}{q}                & All objects that are ever referred to from a variable or field of an object in $q$.\\
    \hoquery{HeapReferredFrom}{q}            & All objects that are ever referred to from a field of an object in $q$.\\
    \hoquery{ReachableFrom}{q}$^\dagger$     & All objects selected by $q$, and all objects that are referred to from fields or variables of objects that are \textq{ReachableFrom}($q$).\\
    \hoquery{HeapReachableFrom}{q}$^\dagger$ & All objects selected by $q$, and all objects that are referred to from fields of objects that are \hoquery{HeapReachableFrom}{q}.\\
    \hoquery{CanReach}{q}$^\dagger$          & All objects selected by $q$, and all objects that held variables or fields pointing at objects that  \hoquery{CanReach}{q}.\\
    \hoquery{CanHeapReach}{q}$^\dagger$      & All objects selected by $q$, and all objects that held fields pointing at objects that \hoquery{CanHeapReach}{q}.\\
    \hoquery{Deeply}{q}$^\dagger$            & All objects in $q$ that \textq{CanReach} only objects also in $q$.\\
    \hoquery{HeapDeeply}{q}$^\dagger$        & All objects in $q$ that \textq{CanHeapReach} only objects also in $q$.\\
    \hoquery{Not}{q}                         & All objects that are not in $q$.\\
    \textq{And($q_1\ \ldots q_N$)}           & All objects that are in all queries.\\
    \textq{Or($q_1\ \ldots q_N$)}            & All objects that are in at least one query.\\
  \end{tabular}
  \caption{Queries and their Meaning. See Section~\ref{sec:queries} for detailed
    descriptions. $^\dagger$these queries are recursively defined, this means
    that these selections effectively ``walk the memory
    graph''.}\label{tab:queries}
\end{table}

\section{Comparisons and Query Refinement}\label{sec:refinement}

We have, so far, shown how to use queries that select objects. When exploring
data sets, it is often useful to interactively compare several queries with each
other and see whether -- or not -- the objects selected by several queries have
large overlaps. Several subqueries, separated by a slash form a composite query:
\lquery{ImmutableObj()/HeapUniqueObj()/TinyObj()}. The result of this query
shows -- amongst other things -- the percentage of
all objects in a particular data set that satisfied each query, 
but also the intersections of the queries, \emph{e.g.,} all objects which are both immutable and tiny. Figure~\ref{fig:matrixFocusHide}(a) shows this information in form of a
matrix.

\subsection{Exploring Selections}
All information that the matrix of a query shows could have been obtained using
the \textq{And} query combinator. However, query compositions are useful because
they form the starting point of exploration of data. A user can execute
operations on either of the subqueries. These operations are exposed by the user
interface as hyperlinks, facilitating speedier interaction with the system.
Being able to modify these queries, in our experience, improves user experience
and the browser's browsing history makes it natural to go back and revisit
queries that a user has seen before.

\subsubsection{Focusing on a Subquery}

To focus on a subquery means to constrain the other subqueries to only select a
subset of the focused query's selection. Given the composite query
\lquery{InstanceOf(java.lang.String)/HeapUniqueObj()}, then to focus on the
query for the heap unique objects would produce the resulting query that selects
all strings that \emph{are also} heap-unique:
\abbrevlquery{And(\-Heap\-Unique\-Obj()
  Instance\-Of(\-java.lang.String))}{And(HeapUniqueObj()\%20InstanceOf(java.lang.String))}.

\begin{figure*}
  \vspace*{-1ex}
  \centering
  \subfloat[No query focused, no query hidden.]{
  \hspace{8mm}\includegraphics[scale=0.8]{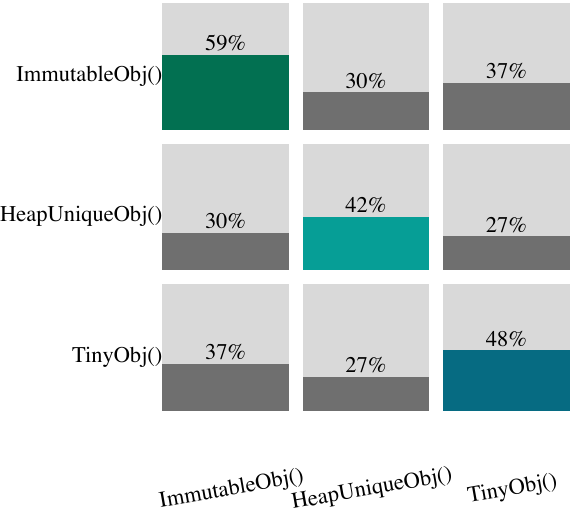}}\label{fig:matrix} 
  \subfloat[After focusing on tiny objects.]{
    \hspace{10mm}\includegraphics[scale=0.8]{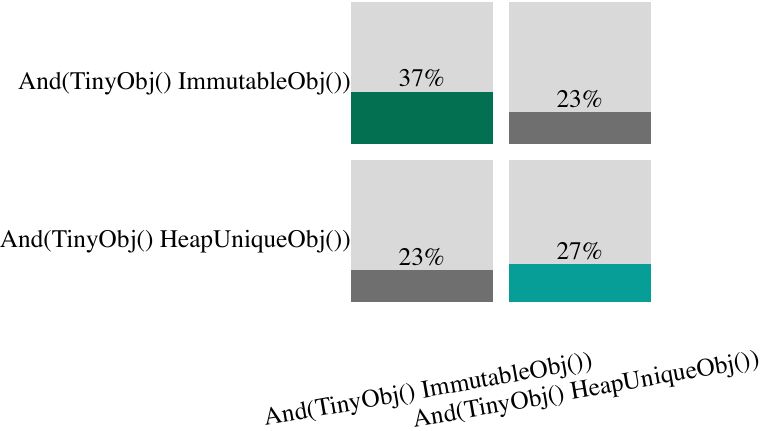}}\label{fig:matrixFocus}\\
  \subfloat[After hiding tiny objects.]{
    \includegraphics[trim=0 0 0 0,scale=0.8,clip]{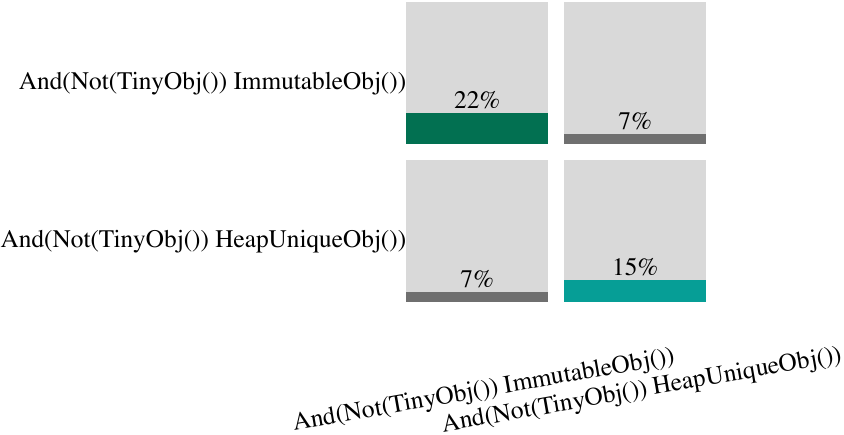}}\label{fig:matrixHide} 
  \subfloat[After splitting by tiny objects. The top-left group of values
  contains tiny objects, the bottom right group contains non-tiny objects. The
  information here is the combination of subfigures (b) and (c).]{
    \includegraphics[trim=0 5 0 0, scale=0.7, clip]{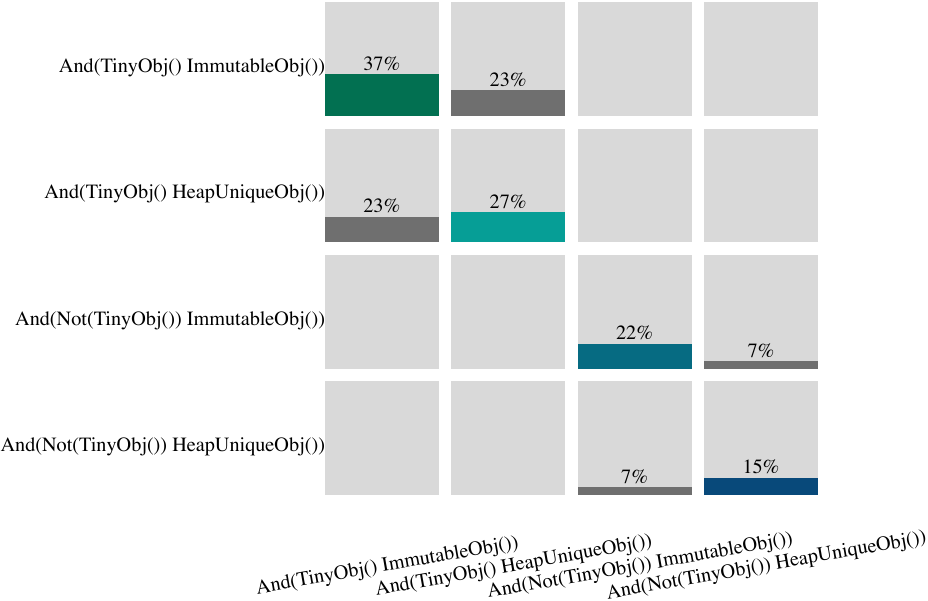}}\label{fig:matrixHide} 
  \caption{Comparing the three queries
    \lquery{ImmutableObj()/HeapUniqueObj()/TinyObj()}, we modify this composite
    query by: focusing (b); hiding (c); or splitting (d)---the tiny objects.}\label{fig:matrixFocusHide}
  \vspace*{-1ex}
\end{figure*}

Figure~\ref{fig:matrixFocusHide}(a) shows a query that consists of three
subqueries, and subfigure (b) illustrates the effects of focusing on one of
them. Table~\ref{tab:refining} depicts the focus operation's transformation of
composite queries.

\begin{table*}
  \begin{tabular}{r@{\code{ / }}r@{\code{ / }}rcl}
    $q_1$\hphantom{)} & $\ldots$ & $q_{N-1}$\hphantom{)} & \code{ / $q_N$} & \\\hline
    \code{And(Not($q_N$) $\ q_1$)} & $\ldots$ & \code{And(Not($q_N$) $\ q_{N-1}$)} & & hiding $q_N$\\
    \code{And($q_N$ $\ q_1$)} & $\ldots$ & \code{And($q_N$ $\ q_{N-1}$)} & & focusing on $q_N$\\
    \code{And($q_N$ $\ q_1$) / And(Not($q_N$) $\ q_1$)} & $\ldots$ & \code{And($q_N$ $\ q_{N-1}$) / And(Not($q_N$) $\ q_{N-1}$)} & & splitting on $q_N$\\
  \end{tabular}
  \caption{Refining a composite query \textq{$q_1$/$\ldots$/$q_n$} by focusing
    or hiding $q_N$.}\label{tab:refining}
\end{table*}


\subsubsection{Hiding a Subquery}

$\!$To hide a subquery means to con\-strain the other subqueries to \emph{never}
select objects in the subquery's selection. It can be thought of set subtraction.
Given the same composite query as before,
\lquery{In\-stance\-Of(java\-.lang\-.String)/Heap\-UniqueObj()}, to hide the
subquery for the heap unique objects would produce the resulting query that
selects all strings that \emph{are not} heap-unique:
\abbrevlquery{And(Not(\-Heap\-Unique\-Obj())
  Instance\-Of(\-java.lang.String))}{And(Not(HeapUniqueObj())\%20InstanceOf(java.lang.String))}.

Hiding a query $q$ is equivalent to negating the query first -- \hoquery{Not}{q}
-- and then focusing on the negated query.

Figure~\ref{fig:matrixFocusHide}(a) shows a query that consists of three
subqueries, and subfigure (c) illustrates the effect of hiding one of them.

\subsubsection{Splitting a Subquery}

For a number of composed queries with a subquery $q_N$, a common question is often
whether the queries $q_1 \ldots q_{N-1}$ yield different results for objects
that are selected by $q_N$ and objects that are not -- a case analysis of sorts.
To \emph{split a subquery $q_N$} means to first, eliminate this query from a
query comparison, and to replace each of the other subqueries $q$ by \emph{two
  new subqueries} \textq{And($q\ q_N$)} and \textq{And($q\ $ Not($q_N$))}, see
Table~\ref{tab:refining}.

To give an example, consider age ordering of objects. An object is age-ordered
if it is younger than all objects it has field references to. An object is
reverse age-ordered if it is older than all objects it has field references to.

Starting with the query comparison
\abbrevlquery{AgeOrderedObj()/Rev\-erseAgeOrderedObj()/InstanceOf(j.l.String)}{AgeOrderedObj()/ReverseAgeOrderedObj()/InstanceOf(java.lang.String)},
and splitting on \lquery{In\-stanc\-eOf(java.lang.String)} gives the resulting
query comparison:

\noindent\abbrevlquery{\small And($\hphantom{\mathtt{Not(}}$InstanceOf(j.l.String)$\hspace{0.8mm}$ AgeOrderedObj()) /\\
  And($\hphantom{\mathtt{Not(}}$InstanceOf(j.l.String)$\hspace{0.8mm}$ ReverseAgeOrderedObj()) /\\
  And(Not(InstanceOf(j.l.String)) AgeOrderedObj()) /\\
  And(Not(InstanceOf(j.l.String))
    ReverseAgeOrderedObj())}{And(InstanceOf(java.lang.String)\%20AgeOrderedObj())/And(InstanceOf(java.lang.String)\%20ReverseAgeOrderedObj())/And(Not(InstanceOf(java.lang.String))\%20AgeOrderedObj())/And(Not(InstanceOf(java.lang.String))\%20ReverseAgeOrderedObj())}\footnote{A
    reader running this composite query might notice that some objects are both
    age ordered and reverse age ordered. The objects that are selected by both
    are tiny objects, which can be verified by filtering out tiny objects,
    which yields the empty set.}

  This comparison makes it evident that strings are more commonly
  reverse age-ordered than other classes.

\section{Case Studies}\label{sec:cases}

This section describes two cases that highlight how Spencer can be
used. Unless otherwise noted, all graphs that are shown here are
renderings that the web user interface produces.

\subsection{Case 1: Exploring the Layout of Strings}\label{sec:case-A}

Our first case study highlights how Spencer can be used to explore a data set
with a potential run-time optimisation in mind. The goal here is not to propose
a run-time optimisation, but to show how a researcher could leverage the
platform in practise. 

Strings (and the character arrays that store their data) are the class with the
highest memory usage in many Java programs \cite{kawachiya2008}. As the page for
the \code{\sourcecodelink{java.lang.String}} class
code\footnote{\url{\sourcecodeurl{java.lang.String}}} shows, Strings in Java are
objects with only one reference type field: the field \code{value} holds a
reference to an array of characters that contains the string's data.
Listing~\ref{lst:string-code} shows an excerpt of the pretty printed Java
bytecode that a user can find on this page.

\begin{lstfloat}
  \vspace*{-1ex}
  \begin{lstlisting}
public final class java/lang/String 
  implements 
    java/io/Serializable 
    java/lang/Comparable
    java/lang/CharSequence {
  // ...
  private final [C value
  private I hash
  // ...
}
\end{lstlisting}
  \vspace*{-1ex}
  \caption{An excerpt of the definition of the class \code{\sourcecodelink{java.lang.String}}, reachable under the view \code{\url{\sourcecodeurl{java.lang.String}}}. \code{[C} in Java bytecode denotes an array of primitive characters and \code{I} denotes a primitive integer.}
  \label{lst:string-code}
\end{lstfloat}

A Spencer query that selects the arrays that are reachable from Strings can be written thus:

A1: \abbrevlquery{\textq{HeapReferredFrom(InstanceOf(j.l.String))}}{HeapReferredFrom(InstanceOf(java.lang.String))}

Strings in Java are, according to the documentation, immutable
\cite{java-lang-string-doc}. Arrays, however, are typically not immutable since
they are constructed empty and must be populated with values. Arrays in strings
could however be \emph{stationary}, meaning that they are fully initialised
before being read. To test whether this intuition is true, we add a new sub
query to compare stationary objects versus the character arrays from before:

\newcommand{\queryvar}[1]{$\langle$#1$\rangle$}

A1.1:
\abbrevlquery{\queryvar{A1}/StationaryObj()}{HeapReferredFrom(InstanceOf(java.lang.String))/StationaryObj()}
\footnote{For clarity we use the abbreviation \queryvar{A1} to
  stand in for its definition above. Actual Spencer query syntax would repeat
  the query A1 there.}

\begin{figure}
  \centering
  \includegraphics[width=8cm]{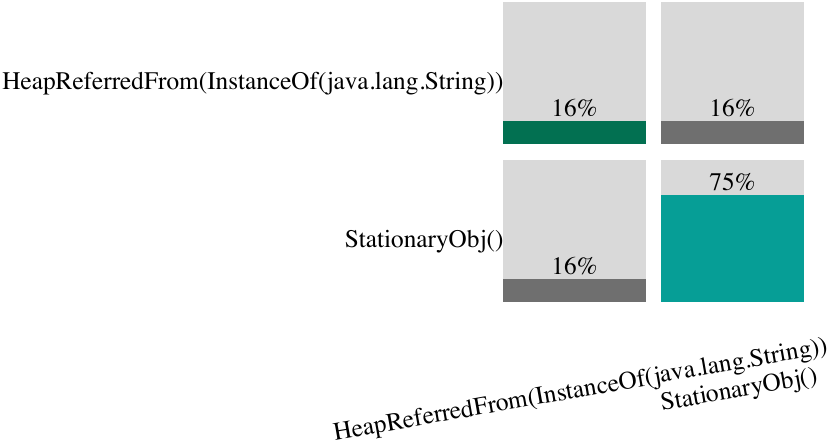}
  \caption{Objects that are heap reachable from strings vs. stationary objects.
    The coloured cells on the main diagonal show the percentage of objects that are
    selected by the query, the objects off the main diagonal show the percentage
    that fulfill both queries.}\label{fig:case1-matrix-stationary}
  \vspace*{-1ex}
\end{figure}

The matrix visualisation of this composite query (shown in
Fig.~\ref{fig:case1-matrix-stationary}) shows that \BLU{75\%} of objects are
stationary, that \BLU{16\%} of objects are heap-referred from strings (top
left), and that also \BLU{16\%} of objects are both heap-referred from strings
and stationary (top right, bottom left). This suggests that all objects referred to from fields in string objects are also stationary -- as a sanity check, we hide
stationary objects and confirm that this selection indeed returns \emph{no} objects:

A1.2: \abbrevlquery{And(Not(StationaryObj()) \queryvar{A1})}{And(Not(StationaryObj())\%20HeapReferredFrom(InstanceOf(java.lang.String)))}

The query A1.2 returns no instances, meaning that these arrays are all
stationary. Thus the the trace data on which these queries are run indicate a
strong possibility of a static invariant that these arrays are stationary. This
might be verifiable from the source code (available directly from Spencer), but
it may not be the case that Java is able to express this, or that the actual
classes involved in an aggregate are known statically.


Stationary data is safe to share from many objects. To detect whether this
fact is leveraged by the string class we can go back to query A1, and add a
subquery to look for objects that have at most one reference from the heap:

A2: \abbrevlquery{\queryvar{A1}/HeapUnique\-Obj())}{HeapReferredFrom(InstanceOf(java.lang.String))\%20/HeapUniqueObj()}

\begin{figure}
  \centering
  \includegraphics[width=8cm]{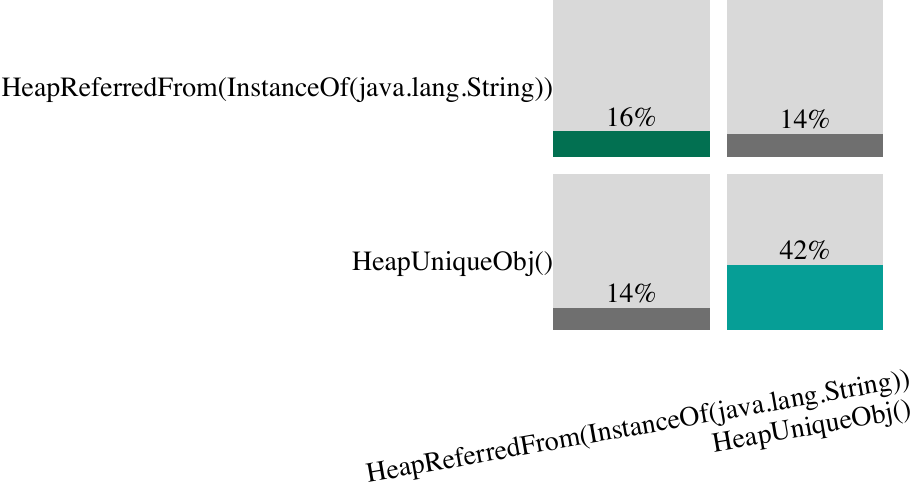}
  \caption{Query A2: Most objects that are heap-referred to from strings are
    heap unique, but not all (\BLU{16\%} are heap-referred to from strings, but
    \BLU{14\%} are both heap-referred to and heap
    unique).}\label{fig:case1-matrix-heapunique}
\end{figure}

The matrix visualisation of this composite query is shown in
Fig.~\ref{fig:case1-matrix-heapunique} and tells us that there are \BLU{14\%} of
objects that are heap reachable from strings \emph{and also} heap unique,
meaning a 1:1 mapping from string objects to character arrays. This means there
are \emph{shared} character arrays in the program, but we do not yet know
whether they are shared by strings. To investigate, we want to focus on the
remaining \BLU{2\%} of objects reachable from strings, but that are not unique:
hiding the heap unique objects does precisely that, yielding the query:

A3: \abbrevlquery{And(Not(HeapUniqueObj())~\queryvar{A1})}{And(Not(HeapUniqueObj())\%20HeapReferredFrom(InstanceOf(java.lang.String)))}

\begin{figure*}
  \centering
  \subfloat[\abbrevlquery{Some query $q$'s selection}{InstanceOf(java.lang.String)}.]{
    \includegraphics[width=0.3\textwidth]{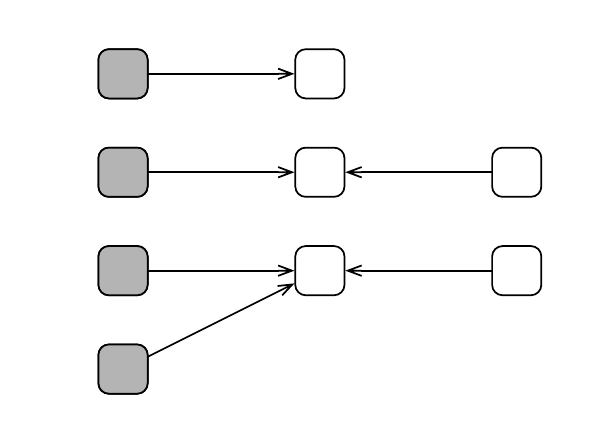}
  }\label{fig:instanceof-string}
  \subfloat[\abbrevlquery{HeapReachableFrom($q$)}{HeapReachableFrom(InstanceOf(java.lang.String)}).]{
    \includegraphics[width=0.3\textwidth]{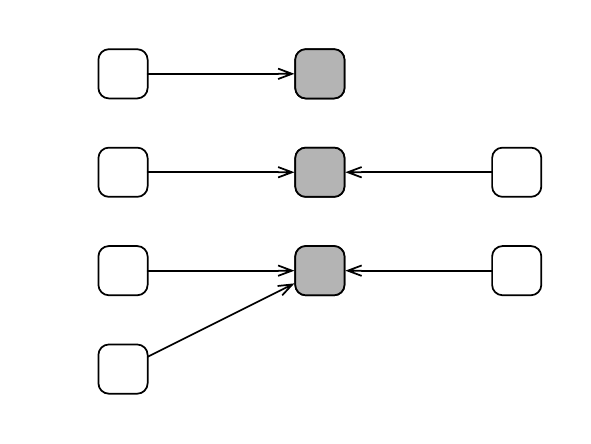}
  }\label{fig:hrf-instanceof-string}
  \subfloat[\abbrevlquery{HeapRefersTo(HeapReachableFrom($q$))}{HeapRefersTo(HeapReachableFrom(InstanceOf(java.lang.String)))}.]{
    \includegraphics[width=0.3\textwidth]{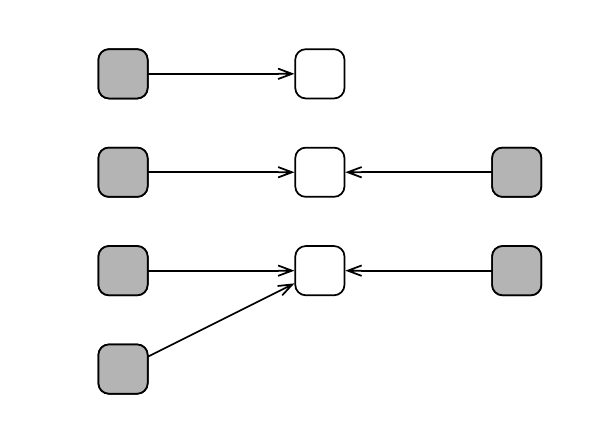}
  }\label{fig:hrt-hrf-instanceof-string}
  \caption{\emph{Gray objects are selected.} \textq{HeapRefersTo} and \textq{HeapReferredFrom} are not inverse
    operations, as this example shows. Case study A in Sec.~\ref{sec:case-A}
    uses such i similar query to find all objects that have a reference to
    non-unique character arrays in strings to find which objects have reference
    to these arrays (these figures were not created using
    Spencer).}\label{fig:case1-query-composition}
\end{figure*}

To see if all of these objects are strings, we select the
objects that hold field references to shared objects by applying the
\textq{HeapRefersTo} query combinator (Fig.~\ref{fig:case1-query-composition}
shows an illustration of what it means to nest a \textq{HeapReferredFrom} query
inside a \textq{HeapRefersTo} query.):

A4: \abbrevlquery{HeapRefersTo(\queryvar{A3})}{HeapRefersTo(And(Not(HeapUniqueObj())\%20HeapReferredFrom(InstanceOf(java.lang.String))))}

To further select the objects (if any) that are not strings, we add a composite
query to select all strings, and hide those:

A5: \abbrevlquery{And(Not(InstanceOf(java.lang.String))
  $\langle{}$A4$\rangle$)}{And(Not(InstanceOf(java.lang.String))\%20HeapRefersTo(And(Not(HeapUniqueObj())\%20HeapReferredFrom(InstanceOf(java.lang.String)))))}

We can inspect the class of the objects that are now selected and find that
these objects are instances of the class
\sourcecodelink{java.lang.StringBuffer}.

If we want to know where those arrays were allocated, we can go back to query
A3, and click inspect on the allocation sites. This reports that all of them
were allocated in file \texttt{StringBuffer.java}, on line 671. inspecting the
source code of the class \code{StringBuffer} (link:
\sourcecodelink{StringBuffer}), we find that this line corresponds to the method
\code{toString()} (see Listing~\ref{lst:stringbuffer-tostring}).

\begin{lstfloat}
  \vspace*{-1ex}
\begin{lstlisting}
  // access flags 0x21
  public synchronized toString()Ljava/lang/String;
   L0
    LINENUMBER 670 L0
    ALOAD 0
    GETFIELD java/lang/StringBuffer.toStringCache : [C
    IFNONNULL L1
   L2
    LINENUMBER 671 L2
    ALOAD 0
    ALOAD 0
    GETFIELD java/lang/StringBuffer.value : [C
    ICONST_0
    ALOAD 0
    GETFIELD java/lang/StringBuffer.count : I
    INVOKESTATIC java/util/Arrays.copyOfRange ([CII)[C
    PUTFIELD java/lang/StringBuffer.toStringCache : [C
   L1
\end{lstlisting}
  \vspace*{-1ex}
  \caption{Instances of the class \code{ShallowImmutable} can never be changed, yet
    it is still not safe to share \code{ShallowImmutable} instances accross threads
    -- the heap-reachable \code{Mutable} instance could cause data
    races.}\label{lst:stringbuffer-tostring}
\end{lstfloat}


\subsection{Case 2: Uncovering Safety Properties of Objects}\label{sec:case-B}

The advent of multicore has renewed the interest in immutability, and caused
mutable state to be criticised. Mutable state that is shared across threads is a
risky programming pattern, and immutable state is suggested as a safe
alternative. A reasonable question then is to investigate how easy it might be
to retrofit abstractions like immutability or uniqueness (references that can
never be aliased) into object-oriented programming as realised in Java. In order
to make such a judgement, we will construct a query that selects all safe
objects, and then analyse the objects that are \emph{not} thread-safe.

We start with immutability:

B1: \lquery{ImmutableObj()}

In our trace data, we find a large number of immutable objects (\BLU{59\%}). Our
notion of immutability is however shallow, and although an object stays the
same, if its aggregate values change, it is arguably not immutable.
Listing~\ref{lst:shallow-immutable} illustrates this.

\begin{lstfloat}
  \vspace*{-1ex}
\begin{lstlisting}
final class ShallowImmutable {
  private final Mutable m;
  public ShallowImmutable(Mutable m) {
    assert(m != null);
    this.m = m;
  }
  public Mutable getMutable() {
    return this.m;
  }
} 
\end{lstlisting}
  \vspace*{-1ex}
  \caption{Instances of the class \code{ShallowImmutable} can never be changed,
    yet it is still not safe to share \code{ShallowImmutable} instances across
    threads -- the heap-reachable \code{Mutable} instance could cause data
    races.}\label{lst:shallow-immutable}
\end{lstfloat}

In order to only select objects whose transitive closure of reachable state is
immutable, we modify the query:

B2: \abbrevlquery{HeapDeeply(\queryvar{B1})}{HeapDeeply(ImmutableObj())}

Instances of the class \code{ShallowImmutable} in
Listing~\ref{lst:shallow-immutable} would not be selected by B2 because
they contain a heap reference to mutable objects of the class \code{Mutable}.
In our dataset, the fraction of deeply immutable objects is \BLU{53\%} -- most objects that are
immutable (B1) are also deeply immutable (B2), only \BLU{6\%} of objects are
immutable but not heap-deeply so.

We can investigate those \BLU{6\%} of objects by constructing a composite query
of B1 and B2, and hiding B2. This yields:

B2.1: \abbrevlquery{And(Not(\queryvar{B2}) \queryvar{B1})}{And(Not(HeapDeeply(ImmutableObj()))\%20ImmutableObj())?vis=klass}

According to the classes of the selected objects, strings are by far the most
common objects of those that are not deeply immutable, but immutable. But in the
previous case study in Sec.~\ref{sec:case-A}, we have learned that strings are
-- even though they do not fulfill the requirements of the \lquery{MutableObj()}
query -- ``morally immutable''. We account for this fact by including them, and
their \code{value} arrays:

B3: \indent\abbrevlquery{Or(InstanceOf(j.l.String)
  \\\indent\hphantom{Or} HeapReferredFrom(\-Instance\-Of(j.l.String)) \\\indent\hphantom{Or} \queryvar{B2}\\\indent)}{Or(InstanceOf(java.lang.String)\%20HeapReferredFrom(InstanceOf(j.l.String))\%20HeapDeeply(ImmutableObj()))}

Stackbound objects (objects that are never referenced from fields) are also
thread safe, as in order to share an object across threads, it needs to pass
through a field at some point (threads can not access each other's stacks
directly). Similarly, unique objects are safe even if they are touched by several
threads -- after all, the objects can not be touched by several threads at the
same time, as there is only one active reference at each time:

B4: \abbrevlquery{Or(StackBoundObj() UniqueObj() \queryvar{B3}))}{Or(StackBoundObj()\%20InstanceOf(java.lang.String)\%20HeapReferredFrom(InstanceOf(j.l.String))\%20HeapDeeply(ImmutableObj()))}

In programming language design, a possible pitfall is to design abstractions
that fit simple cases well but that are not able to support real world use
cases. Imagining we're implementing a type system for a Java-like language that
has type abstractions for stack-boundedness, uniqueness, and immutability (there are many such works in the literature \emph{e.g,.} \cite{gordon, burying, boyland2001, Nelson2013, stationary-inference, otsurvey}).
We would like to understand what are the objects that are not ``safe'', -- to see the potential usefulness of our type system, and also understand the objects
that are unlikely to fit our abstractions:

B5: \abbrevlquery{Not(\queryvar{B4})}{Not(Or(StackBoundObj()\%20InstanceOf(java.lang.String)\%20HeapReferredFrom(InstanceOf(j.l.String))\%20HeapDeeply(ImmutableObj())))}

\begin{figure}
  \centering
  \includegraphics[trim=40 275 31 80, width=8cm, clip]{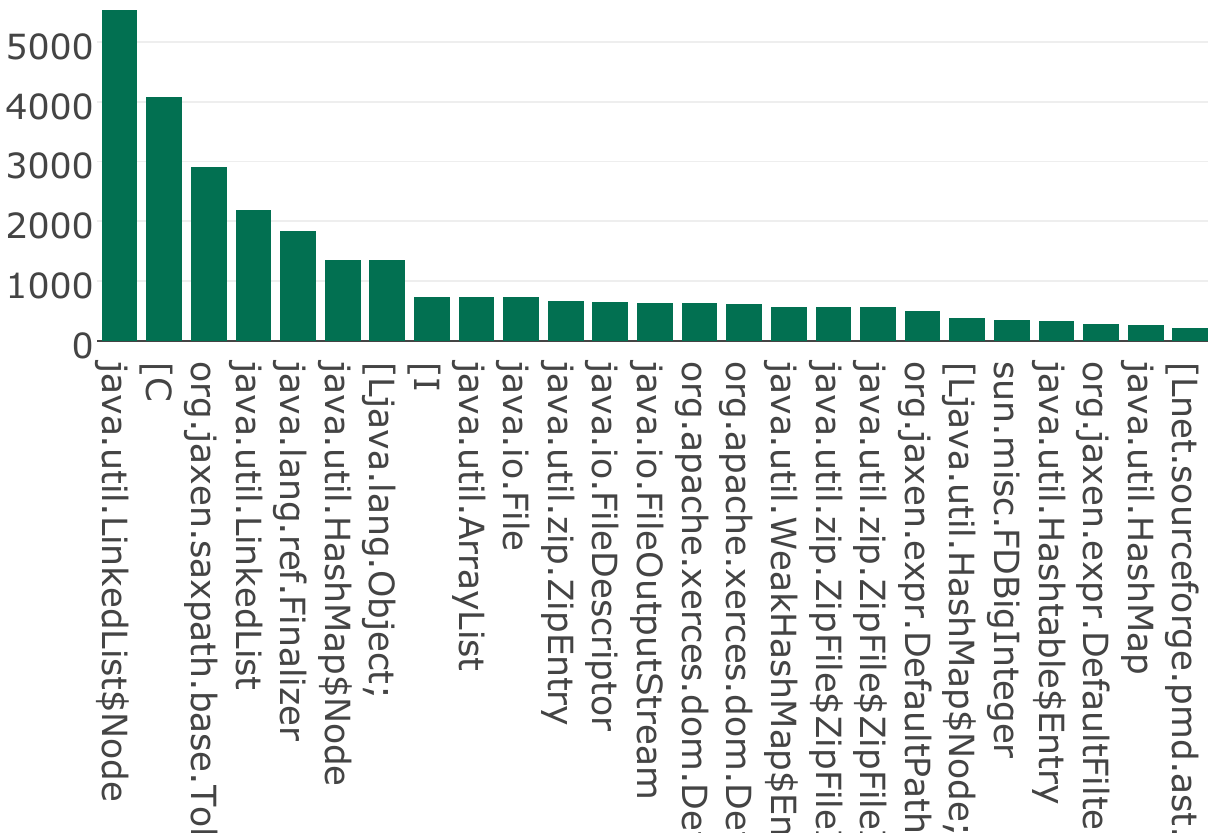}
  \caption{Query B5: Classes of objects that are ``unsafe'', a programming
    language design that deals with the abstractions in case study B
    (Sec.~\ref{sec:case-B}) must be careful to be able to express these
    classes.}\label{fig:unsafe-barchart}
\end{figure}

Looking at the classes of these ``unsafe objects'', we see the bar chart in
Fig.~\ref{fig:unsafe-barchart}. It tells us, for instance, that Nodes of linked
data structures are problematic for such a type system design: the class
\href{\sourcecodeurl{java.util.LinkedList\%24Node}}{\code{java.util.LinkedList\$Node}}
is the most commonly used unsafe class. This result is correct: nodes are aliased from the heap and nodes are referenced both
from the previous \emph{and} from the following node (linked lists in Java are doubly
linked \cite{java-util-linkedlist-doc}). Nodes are also mutable, as
building a list requires changing the \code{next} field of the nodes. And, since
it is a linked data structure, these objects are not stackbound either. Were we
developing such a type system, we would now have identified a possible
shortcoming that we would need to address, given how central linked lists are to many programs.

\section{Internal Details}

Spencer, the web based tool requires a tool chain behind the scenes to function.
This tool chain includes three key programs:

\newcommand{\spencertrace}{\code{spencer-trace}\xspace}
\newcommand{\spencerload}{\code{spencer-load}\xspace}
\newcommand{\hotspot}{HotSpot\textsuperscript{TM}\xspace}

\begin{enumerate}\raggedright
\item A tool, called \spencertrace, to modify all code loaded in a running Java
  virtual machine to emit event logs (see Sec.~\ref{sec:spencer-trace}).
\item A tool, called \spencerload, to load these event logs into a data
  base (see Sec.~\ref{sec:spencer-load}).
\item Spencer, the web application that has been presented above.
\end{enumerate}

\subsection{Tracing with \code{spencer-trace}}\label{sec:spencer-trace}

The \spencertrace tool is a wrapper for the Java \hotspot VM. It is intended to
understand all arguments that \hotspot understands and therefore to serve as a
drop-in replacement for it. When \spencertrace runs a compiled program,
additionally to executing Java bytecode, it will intercept loading of any
bytecode (whether from disk, or dynamically generated, or via other sources),
and transform the loaded code. The implementation of this is backed by a JVMTI
(JVM Tool
Interface\footnote{\url{http://docs.oracle.com/javase/8/docs/platform/jvmti/jvmti.html}})
agent. JVMTI makes it possible to intercept loading of classes by implementing a
handler. In this handler implementation, \spencertrace sends the code of the
class to a code transformation library that modifies the code as shown in
Listing~\ref{lst:code-instrumentation}. The transformation inserts calls to
methods into the code that will write events in a standardised format to disk.
The tool takes care to not instrument data that are used during instrumentation
(like the code that instruments classes itself) by doing the transformation and
logging in native (C-code) implementations and by running the class
transformation in a separate JVM process. Listing~\ref{lst:code-instrumentation}
shows a description of the inserted instrumentation. From this listing, it is
easy to see that the overhead the tracing incurs is substantial. This is a
problem for programs who have built-in time outs, but our experience with
generating trace data is that albeit slow, it works solidly. The fact that the
data from one trace can be used to perform many analyses also mitigates the
slowdown.

Additionally, \spencertrace stores both the original and the transformed version
of the class file into a log directory. The transformed version of the class
file permits running programs with only some classes of interest being
instrumented which boosts performance considerably.

One issue \spencertrace faces is actions taken by methods implemented natively
-- these methods not be instrumented currently. One such method is
\code{System.arraycopy}, a native implementation that copies a range from one
array into another array. One effect this has is that character arrays of
strings often will appear to be immutable, when in fact they were not -- because
the writes that \code{arraycopy} generates are missing in the data. We are 
considering wrapping the most common  native methods to overcome this problem.

\begin{lstfloat}
  \begin{lstlisting}
public int hashCode() {
+ try {
+   NI.methodEnter(.."hashCode"..);
+   NI.read(.."hash"..);
    int h = this.hash;
    
+   NI.loadFieldA(..this, "value",
                    this.value,..);
    if (h == 0 && this.value.length > 0) {
+     NI.storeVar(..2, this.value..);
      char val[] = this.value;

+     NI.loadFieldA(..this, "value"..);
      for (int i=0;
           i<this.value.length;
           i++) {
+       NI.loadArray(..val, i, this..);
        h = 31 * h + val[i];
      }
+     NI.modify(..this, "hash"..);
      this.hash = h;
    }
+   NI.methodExit(.."hashCode"..)
    return h;
+ } catch (Throwable t) {
+   NI.methodExit(.."hashCode"..);
+   throw t;
+ }
}
  \end{lstlisting}
  \caption{The instrumentation adds calls to record what a method was doing.
    This listing shows the effect of code transformation on the method
    \code{hashCode} of the class \code{java.lang.String}. The transformation
    works on Java bytecode, not Java code, this presentation in Java is
    therefore just for illustration. Lines that start with \code{+} are added by
    the transformation. Variables of reference type (classes, arrays) are
    instrumented, primitive variables are currently not. The
    \code{NativeInterface} interface (abbreviated as \code{NI}) is added by
    \spencertrace, the method implementations write the data to
    disk.}\label{lst:code-instrumentation}
\end{lstfloat}

\newcommand{\capnproto}{Cap'n Proto\xspace}

The tracefiles are written in a standardised format, using a specification
compatible with the \capnproto tool. \capnproto accepts record-like
specifications as input and generates libraries in a number of languages (the
supported languages include C, C++, Java, C\#, Go, OCaml, Ruby, Javascript, and
others) that can be used to write these records to disk or read them back. Using
this approach, users could implement their own tools that generate trace files
(perhaps for languages that are not running on the JVM at all) and spencer could
then host these traces just as well.

\begin{lstfloat}
  \centering
  \vspace*{-1ex}
  \begin{lstlisting}
      struct VarStoreEvt {
        callermethod  @0 :Text;
        callerclass   @1 :Text;
        callertag     @2 :Int64;
        newval        @3 :Int64;
        oldval        @4 :Int64;
        var           @5 :Int8;
        threadName    @6 :Text;
      }
  \end{lstlisting}
  \vspace*{-1ex}
  \caption{A specification for an event that represents a variable assignment in
    the \capnproto's input language. \capnproto can use such specifications to
    generate optimised libraries in a range of languages that will write these
    events to disk or read them back into memory.}\label{lst:capnp-spec}
\end{lstfloat}

\subsection{Loading Traces with \spencerload}\label{sec:spencer-load}

Program traces can be loaded into a local database using the \spencerload tool.
The tool reads the logs from a file produced by \spencertrace
(Sec.~\ref{sec:spencer-trace}) and loads them into a PostgreSQL database.

This database contains tables to track: objects (identified by a unique ID)
containing the object's class, the event number of their first appearance in the
trace, and the event number of their last appearance; references between
objects (including when the reference was established; whether stored in a variable
or in a field; when the reference ended), and method calls (including
which object was being called, name and signature of the method, when the method
was called, and when the call returned).

\subsection{Available Data}

Spencer currently hosts traces of 9 of the programs in the DaCapo program corpus
version 9.12 \cite{dacapo} (\emph{c.f.}
\url{http://www.spencer-t.racing/datasets}) comprising more than 3 billion
events and over 2.000 loaded classes including libraries. The selection of
programs currently only features Java programs but will eventually grow to
include programs in other languages (like Scala, Clojure, and Ruby), too. The
number of current programs is limited by the hard drive capacity that the
current server has and we are working on increasing the available space and
their size is limited by available RAM\footnote{To reviewers: The current server
  provides only 16GB of RAM, which is not sufficient to run some queries on the
  larger benchmarks like \dataset{avrora} and \dataset{h2}, but we hope that the
  amount of RAM will be at least doubled in the next few weeks. Additionally, we
  will implement optimisations that conserve memory.}. In principle, Spencer can
instrument any application running on the JVM. A common problem, however, is
asynchronously communicating software that puts bounds on reply latencies. The
slowdown of Spencer can, in such cases, cause time-outs to happen. Such software
may -- where possible -- need to be run with changed parameters in order to
account for the slowdown.

\begin{table}
  \centering
  \begin{tabular}{lrr}
    Name               & Objects    & Log            \\\hline
    \dataset{luindex}  & 81,158     & 5.8GB          \\
    \dataset{pmd}	     & 131,462    & 2.7GB          \\
    \dataset{fop}      & 521,789    & 10GB           \\
    \dataset{batik}    & 526,945    & 21GB           \\
    \dataset{xalan}    & 1,133,391  & 48GB           \\
    \dataset{lusearch} & 1,212,743  & 61GB           \\
    \dataset{sunflow}  & 2,419,900  & 91GB           \\
    \dataset{h2}       & 6,655,852  & 207GB          \\
    \dataset{avrora}   & 932,085    & 236GB          \\\hline\hline
    Total:             & 13,615,325 & $\approx$680GB \\
  \end{tabular}
  \caption{Currently loaded benchmarks, a similar list can be found in the tool:
    \url{\spencerurl/datasets}.}\label{tbl:datasets}
\end{table}

\newcommand{\roadrunner}{\textsc{RoadRunner}\xspace}

\section{Related Work}

Many tools for dynamic analysis have been developed in the past. The JVM,
historically has been a good basis, due to tools like JVMTI that we also rely
on.

What sets Spencer apart from the previous work is that Spencer is -- mostly -- a
collection of data with tools to analyse them in ways that makes it easy to
explore data, unearth knowledge about a running program, and share the results
or collaborate. Previous work on dynamic tracing focus on collecting the data.
In this regard, Spencer contributes the inclusion of variable events in traces,
which previous state-of-the-art tools like \roadrunner{} \cite{roadrunner} does not
do.

\subsection{Snapshotting for Heap Analysis}

Snapshotting is a sampling-based dynamic analysis technique that regularly stops a running
program and writes all contents of the heap to disk. The snapshots can then be
analysed offline.

The advantages of snapshotting include that the amount of data generated
generally is lower; the disadvantages include that it is unknown what happened
to an object in between snapshots. Having continuous data about an object's
execution permits Spencer to provide convincing results in cases where snapshots
would not be able to do this: if many object have only one incoming reference in
a heap snapshot, that does not mean that unique variables are a useful type
abstraction -- because they might be aliased before and after the snapshot.
Temporary violations of uniqueness, or ABA-style updates to objects will be
invariable by caught by spencer, and this is part of the implementation of uniqueness and immutability.
An advantage of snapshotting is its ability to deal with native code. 

Potanin et. al's Fox \cite{Potanin2004} relies on snapshotting and also uses a
query language. Potanin et. al used Fox to look for uniqueness in the heap of
Java programs. The proportion of aliased objects found in their corpus was
13.6\% on average. This is roughly in line with values that we observe for the
query \lquery{Not(HeapUnique()} but, as stated above, the measuring methodology
differs as Spencer is able to track all events on an object. While the results
are similar, it is unclear whether the objects reported as \emph{e.g.,} unique
are the same across both tools.


Interestingly, the case study about strings in Section~\ref{sec:case-A} is
similar to approaches used in real world development of programming language
run-times: heap snapshots (also called heap dumps) have been used to gain
insights into run-time optimisations -- before implementing. One example is work
by Oracle on compressed strings in Java (strings can only use one byte per
character, rather than two, in case they only contain ASCII characters). There,
heap dumps of strings have been used to estimate how much application memory
could be saved by implementing such an optimisation (\emph{``We have a large
  corpus of heap dumps [..] that can be used to estimate the heap occupancies
  for some frequent classes. So, we have crafted a simple simulator that
  introspected what Strings are there, [..] and how compressible those
  characters are.''} \cite{Shipilev2016})\footnote{In general, it is hard to
  translate application memory savings to virtual memory savings, as JVMs use
  sophisticated memory optimisations, that depend on runtime settings. We do not
  know of runtime optimisations that can automatically reduce the size of
  arrays, so savings obtained by using smaller arrays are likely to translate
  directly to virtual memory savings.}. Spencer can, right now, not deal with
this case, as we do not store what primitive values are assigned into arrays. As
noted in the section on future work (Sec.~\ref{sec:future-work}), we plan to
extend tracing of primitive values, and also record sizes of arrays to support
such use cases.

Chis et al. analyse heap snapshots, focusing on memory bloat in Java programs
and identifies common problems that are specific to Java programs
\cite{patterns-inefficency}. These problems include implementation-specific
issues: for example, they find many empty \code{java.util.ConcurrentHashMap}
objects that contain few or no elements but use considerable amounts of memory
each and suggest practical fixes to get rid of this overhead. Mitchell et
al.~\cite{runtime-structure-object-ownership}, summarises heap snapshots in ways
that programmers may comprehend with a different goal than ours---to identify
memory bloat. 

\subsection{Trace-Based Tools}

The \roadrunner tracing framework \cite{roadrunner} is a state-of-the-art
tracing tool developed by Flanagan et. al. The goal of \roadrunner to simplify
obtaining trace data by facilitating rapid prototyping of dynamic analyses for
concurrent Java programs. \roadrunner is in this respect similar to the \texttt{spencer-tracing} tool, but does not have any counterparts to the remaining Spencer tool-chain. 
\roadrunner and \texttt{spencer-tracing} export a similar feature set, but the latter importantly include variable events in its traces. 

One similarity that Spencer and \roadrunner have is extensibility: \roadrunner
supports combining several analyses using analyses as a ``chain of filters''
where each filter reduces the number of objects that the subsequent filter has
to analyse. Spencers query combinators are more powerful than that, one example
being the ability to walk the heap using query combinators like
\textq{CanReach($q$)}.


\subsection{Static Analysis}

Hackett and Aiken use static analysis on C programs \cite{aliasing-systems-sw}.
They are able to uncover some static invariants of C programs, but are not able
to gauge \emph{e.g.,} how common certain kinds of objects are at run-time.
Following their static analysis approach is less feasible in Java programs
because of the additional problems that must be solved, such as dealing with
dynamic dispatch and dynamic code generation.

Unkel and Lam \cite{stationary-inference} use static analysis on Java benchmarks
and open source programs to detect the number of stationary fields. Nelson et.
al later study the same property using dynamic analysis \cite{Nelson2013}. They
find the number of stationary fields to be in the range of 55--82\% in a variety
of programs (the static analysis giving the lower bound and the dynamic analysis
giving the upper bound). Spencer measures a stronger property -- stationary
\emph{objects}, which are objects with only stationary fields.

Vanciu et al. \cite{vanciu} use static analysis on hand-annotated programs to
extract Object Ownership Graphs (OOG). It is a conservative whole-program
analysis that shows all possible objects and all possible communication between
objects. These graphs do not scale well to large programs.

\section{Future Work}\label{sec:future-work}

Future work on Spencer can be divided into four main categories. First,
extending the data sets with more programs and more traces of single programs
with varying inputs. Second, the Spencer feature set will be extended by more
queries, including thread-locality, access patterns, etc. Third, we will use
Spencer to validate designs, both our own, and those of others. For example,
there exist many proposals (\emph{e.g., }\cite{boyland2001, EU, gordon, otsurvey}) for type
system designs to rule out certain classes of errors that include unique
references, immutable objects, etc. It would be interesting to see the extent to
which such systems could describe the shapes of existing programs.
%
%
Fourth, improsed user interface, improved visualisation and object interaction. 
For example, a currently missing feature is the ability to select
objects from the visualisations. For example, we are often interested in the
outliers of a statistic -- where are the objects that live the longest
allocated? What are their classes? To answer such questions, one must download
Spencer data dumps and write one's own analyser. This is suboptimal.

\section{Conclusion}

We have presented Spencer, a web based tool for easy, reproducible heap analysis
for programs running on the JVM. Spencer, we believe, will be useful for
researchers in the field of programming languages like ourselves, who want input
to design decisions, and rule out ideas on an early stage, and for researchers
who wish to understand and quantify Java heaps. What sets Spencer aside from
previous work is its aim to simplify exploration of a particular kind of data
set to answer a more narrow set of questions, rather than provide a tracing
solution that ``fits all''. This limits its usefulness as a general-purpose
tool, but greatly simplifies the user experience.

The nine programs that make up the Spencer tool-chain comprise approximately
10,000 LOC in a mixture of Java, C++, and Scala. They are all open-source and
available on GitHub, but more imprortantly Spencer is provided as a free service
hosted by Uppsala University. The continously growing data set currently weights
in at 680GB.

~

\bibliographystyle{plain}
\bibliography{main}

\begin{thebibliography}{10}

\bibitem{java-lang-string-doc}
{Java Documentation: java.lang.String}.
\newblock \url{http://docs.oracle.com/javase/8/docs/api/java/lang/String.html}.
\newblock Accessed: 2017-02-09.

\bibitem{java-util-linkedlist-doc}
{Java Documentation: java.util.LinkedList}.
\newblock
  \url{http://docs.oracle.com/javase/7/docs/api/java/util/LinkedList.html}.
\newblock Accessed: 2017-02-09.

\bibitem{Shipilev2016}
{Q\&A with Aleksey Shipilev on Compact Strings Optimization in OpenJDK 9}.
\newblock \url{https://www.infoq.com/news/2016/02/compact-strings-Java-JDK9}.
\newblock Accessed: 2017-02-08.

\bibitem{dacapo}
S.~M. Blackburn, R.~Garner, C.~Hoffman, A.~M. Khan, K.~S. McKinley, R.~Bentzur,
  A.~Diwan, D.~Feinberg, D.~Frampton, S.~Z. Guyer, M.~Hirzel, A.~Hosking,
  M.~Jump, H.~Lee, J.~E.~B. Moss, A.~Phansalkar, D.~Stefanovi\'{c},
  T.~{VanDrunen}, D.~von Dincklage, and B.~Wiedermann.
\newblock {The DaCapo Benchmarks: Java Benchmarking Development and Analysis}.
\newblock In {\em OOPSLA '06: Proceedings of the 21st annual ACM SIGPLAN
  conference on Object-Oriented Programing, Systems, Languages, and
  Applications}, pages 169--190, New York, NY, USA, October 2006. ACM Press.

\bibitem{burying}
John Boyland.
\newblock Alias burying: Unique variables without destructive reads.
\newblock {\em Softw., Pract. Exper.}, 31(6):533--553, 2001.

\bibitem{boyland2001}
John Boyland, James Noble, and William Retert.
\newblock Capabilities for sharing.
\newblock In {\em ECOOP 2001—Object-Oriented Programming}, pages 2--27.
  Springer, 2001.

\bibitem{patterns-inefficency}
Adriana~E. Chis, Nick Mitchell, Edith Schonberg, Gary Sevitsky, Patrick
  O'Sullivan, Trevor Parsons, and John Murphy.
\newblock {Patterns of Memory Inefficiency}.
\newblock {\em Lecture Notes in Computer Science (including subseries Lecture
  Notes in Artificial Intelligence and Lecture Notes in Bioinformatics)}, 6813
  LNCS:383--407, 2011.

\bibitem{EU}
Dave Clarke and Tobias Wrigstad.
\newblock {External Uniqueness Is Unique Enough}.
\newblock In Luca Cardelli, editor, {\em ECOOP 2003 – Object-Oriented
  Programming}, volume 2743 of {\em Lecture Notes in Computer Science}, pages
  176--200. Springer Berlin Heidelberg, 2003.

\bibitem{otsurvey}
Dave Clarke, Johan Östlund, Ilya Sergey, and Tobias Wrigstad.
\newblock {Ownership Types: A Survey}.
\newblock In Dave Clarke, James Noble, and Tobias Wrigstad, editors, {\em
  Aliasing in Object-Oriented Programming. Types, Analysis and Verification},
  volume 7850 of {\em Lecture Notes in Computer Science}, pages 15--58.
  Springer Berlin Heidelberg, 2013.

\bibitem{roadrunner}
Cormac Flanagan and Stephen~N. Freund.
\newblock The roadrunner dynamic analysis framework for concurrent programs.
\newblock In {\em Proceedings of the 9th ACM SIGPLAN-SIGSOFT Workshop on
  Program Analysis for Software Tools and Engineering}, PASTE '10, pages 1--8,
  New York, NY, USA, 2010. ACM.

\bibitem{gordon}
Colin~S. Gordon, Matthew~J. Parkinson, Jared Parsons, Aleks Bromfield, and Joe
  Duffy.
\newblock {Uniqueness and Reference Immutability for Safe Parallelism}.
\newblock {\em SIGPLAN Not.}, 47(10):21--40, October 2012.

\bibitem{aliasing-systems-sw}
Brian Hackett and Alex Aiken.
\newblock {How is Aliasing Used in Systems Software?}
\newblock In {\em Proceedings of the 14th ACM SIGSOFT international symposium
  on Foundations of software engineering}, SIGSOFT '06/FSE-14, pages 69--80,
  New York, NY, USA, 2006. ACM.

\bibitem{haller2010}
Philipp Haller and Martin Odersky.
\newblock {Capabilities for Uniqueness and Borrowing}.
\newblock {\em 24th European Conference on Object-Oriented Programming (ECOOP
  2010)}, (June):354--378, 2010.

\bibitem{kawachiya2008}
Kiyokuni Kawachiya, Kazunori Ogata, and Tamiya Onodera.
\newblock Analysis and reduction of memory inefficiencies in java strings.
\newblock {\em SIGPLAN Not.}, 43(10):385--402, October 2008.

\bibitem{runtime-structure-object-ownership}
Nick Mitchell.
\newblock {The Runtime Structure of Object Ownership}.
\newblock {\em ECOOP 2006–Object-Oriented Programming}, pages 74--98, 2006.

\bibitem{Nelson2013}
S~Nelson, D~J Pearce, and J~Noble.
\newblock {Profiling Field Initialisation in Java}.
\newblock {\em Lecture Notes in Computer Science (including subseries Lecture
  Notes in Artificial Intelligence and Lecture Notes in Bioinformatics)}, 7687
  LNCS:292--307, 2013.

\bibitem{Potanin2004}
Alex Potanin, James Noble, and Robert Biddle.
\newblock {Checking ownership and confinement}.
\newblock {\em Concurrency Computation Practice and Experience},
  16(7):671--687, 2004.

\bibitem{stationary-inference}
Christopher Unkel and Monica~S. Lam.
\newblock {Automatic Inference of Stationary Fields: A Generalization of Java's
  Final Fields}.
\newblock In {\em Proceedings of the 35th Annual ACM SIGPLAN-SIGACT Symposium
  on Principles of Programming Languages}, POPL '08, pages 183--195, New York,
  NY, USA, 2008. ACM.

\bibitem{vanciu}
Radu Vanciu and Marwan Abi-Antoun.
\newblock {Object graphs with ownership domains: An empirical study}.
\newblock {\em Lecture Notes in Computer Science (including subseries Lecture
  Notes in Artificial Intelligence and Lecture Notes in Bioinformatics)},
  7850:109--155, 2013.

\bibitem{westbrook2012}
Edwin Westbrook, Jisheng Zhao, Zoran Budimli, and Vivek Sarkar.
\newblock Practical permissions for race-free parallelism.
\newblock In James Noble, editor, {\em ECOOP 2012}, volume 7313 of {\em LNCS},
  pages 614--639. Springer, 2012.

\end{thebibliography}

\end{document}